\documentstyle[epsfig,longtable]{aipproc}

\begin{document}

\title{Gravitational Lensing and the Hubble Deep Field}

\author{Asantha R. Cooray, Jean M. Quashnock, M. Coleman Miller}
\address{Dept. of Astronomy and Astrophysics, University of Chicago, 
Chicago IL 60637. \\ E-mail: asante@hyde.uchicago.edu}

\maketitle

\begin{abstract}

We calculate the expected number of multiply-imaged
galaxies in the Hubble Deep Field (HDF), using photometric redshift
information for galaxies with $m_I < 27$ that were detected
in all four HDF passbands.
A comparison of these expectations with
the observed number of strongly lensed galaxies constrains
the current value of $\Omega_m-\Omega_{\Lambda}$, where $\Omega_m$ is
the mean mass density of the universe and $\Omega_\Lambda$ is
the normalized cosmological constant. Based on current estimates of the
HDF luminosity function and associated uncertainties in individual
parameters, our 95\% confidence lower limit on
$\Omega_m-\Omega_{\Lambda}$ ranges between -0.44, if there are no strongly
lensed galaxies in the HDF, and -0.73, if there are two strongly
lensed galaxies in the HDF. 
If the only lensed galaxy in the HDF is the one presently
viable candidate, then, in a flat universe
($\Omega_m+\Omega_\Lambda=1$), $\Omega_{\Lambda} < 0.79$ (95\% C.L.).
These limits are compatible with estimates based on high-redshift
supernovae and with previous limits based on gravitational lensing.

\end{abstract}

\section*{Introduction}

The Hubble Deep Field (HDF; [1]) is the deepest optical survey
that has been performed to date, allowing detailed studies of the
galaxy redshift distribution and the global star formation history.
Galaxies in the HDF have redshifts which are estimated to range from 0.1
to 5, with a large portion having redshifts between 2 and 4.
Such galaxies have a significant probability of being strongly lensed.

The combination  of high resolution and deep exposures in multiple colors
provides a rich ground for gravitational lens searches,
and it was expected that the 
HDF would contain between 3 to 10 lensed galaxies,
based on the number of lensed quasars and radio sources in other surveys [2]. 
Instead, a careful analysis of the HDF [3] has
revealed a surprising dearth of candidates for lensed sources. In fact,
the best estimate is either 0 or 1 lensed sources in the entire field,  
although very faint images with small angular  separations 
may have escaped current analyses.  
This lack of lensing has led to suggestions (e.g., Ref. [3])
that the HDF data may be incompatible with the high probability
of lensing expected in a universe with a large cosmological constant.

Here, we calculate the expected number of detectable,
multiply-imaged galaxies in the HDF
for different cosmological parameters, and we constrain
these parameters by comparing the expectations with the observations.
Further details of our calculation can be found in Ref. [4].

\section*{Expected Number of Lensed Galaxies}

In order to calculate the number of lensed galaxies in the HDF, we model the
lensing galaxies as singular isothermal spheres (SIS) and use the
analytical  filled-beam approximation (e.g., Ref. [5]).
We also include the effects of ``magnification bias'' due to
a magnitude limit of point-source detection in the I-band.
To describe the foreground lensing galaxies, we assume that the
brightness distribution of galaxies
at any given redshift is described by a Schechter function.
We use the redshift-dependent luminosity function given in Ref. [6]
to describe the foreground galaxies.

The background galaxies are described by magnitude and
redshift distributions given in two photometric redshift catalogs [6,7]. 
These catalogs are complete to an I-band limiting magnitude of 27 
and contain a total of 848 galaxies.
In Figure~1,
we show the estimated photometric redshifts of the HDF galaxies according to
Ref. [6] ({\em left panel}) and Ref. [7] ({\em right panel})
versus I-band magnitude. Even though the
redshift distribution is different for the two catalogs ---
because redshifts are estimated using two different techniques ---
our constraints on cosmological parameters are almost the same for
either catalog.

\begin{figure}
\centerline{\epsfig{file=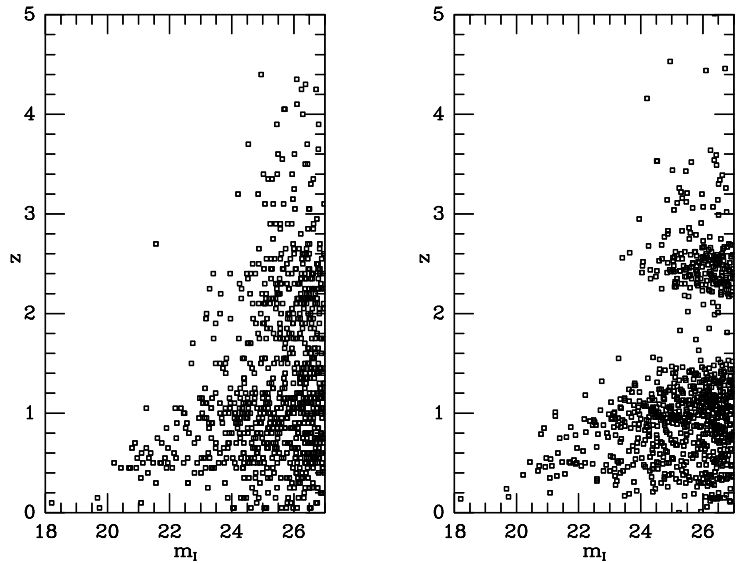,width=2.9truein,angle=-90}}
\vspace{10pt}
\caption{Redshift-magnitude distribution of 848 galaxies with
I-band magnitude $< 27$ in the HDF. The plot shows the estimated
photometric redshifts in the catalogs of Ref. [6] ({\em left panel}) 
and Ref. [7] ({\em right panel}) versus I-band magnitude.
Both catalogs appear to trace the same redshift distribution, with two
peaks ($z \sim 0.6$ and 2.3). However, there is a lack of galaxies in the
Ref. [7] catalog between $z \sim 1.5$ to 2.2. This is the same range
in redshift where no spectroscopic redshifts are currently available for the
HDF.}
\end{figure}

\section*{Constraints on Cosmological Parameters}

In Figure~2, we show the expected number, $\bar N$, of strongly lensed
sources in the HDF as a function of $\Omega_m$ and $\Omega_\Lambda$,
using the photometric redshift catalog of Ref. [6].
A universe dominated with $\Omega_\Lambda$ has a higher number of 
multiply-imaged
sources than the number in a universe dominated with a large $\Omega_m$.
As shown in Fig.\ 2,
$\bar N$ is essentially a function of the combined quantity
$\Omega_m-\Omega_{\Lambda}$. This degeneracy in the lensing probability
permits us to constrain $\Omega_m-\Omega_{\Lambda}$
rather than $\Omega_m$ or $\Omega_{\Lambda}$ separately.

We constrain the quantity $\Omega_m - \Omega_\Lambda$
by comparing the observed and expected number of
lensed galaxies in the HDF and by using a Bayesian likelihood approach [4].
We consider cases in which either 0, 1 (the best estimate from observations), 
or 2 lensed sources are found in the HDF,
for lens search programs that have been carried out 
to an I-band limiting magnitude of $m_{\rm lim}=28.5$.

If there are no lensed galaxies in the HDF, 
then at the 95\% confidence level $\Omega_m-\Omega_{\Lambda} > -0.44$, 
so that in a flat universe  $\Omega_{\Lambda} < 0.72$. 
If there is one lensed galaxy in the HDF, 
our constraint depends only slightly on the galaxy redshift, 
estimated to be between 1 and 2.5 [3].
If the galaxy redshift is 1, 
then $\Omega_m-\Omega_{\Lambda} > -0.52 $, 
implying $\Omega_{\Lambda} < 0.76$ in a flat universe. 
If instead the galaxy redshift is 2.5, 
then  $\Omega_m-\Omega_{\Lambda} > -0.58$, 
and hence $\Omega_{\Lambda} < 0.79$ in a flat universe.
If 2 strongly lensed galaxies are present in the HDF, 
$\Omega_m-\Omega_\Lambda > -0.73$ at the 95\% confidence level. 
These limits are compatible with estimates based on high-redshift
type Ia supernovae [8]
and with previous limits on the cosmological constant
based on gravitational lensing [9,10].

In Ref. [4], we discussed some of the systematic uncertainties in
our calculation, the largest of which are due to effects of
reddening and extinction on (optical) lens search programs. 
Such effects tend to reduce the number of observed lenses,
and hence could lead to a systematic underestimate
of the upper bound on the cosmological constant.
We allowed $m_{\rm lim}$ to vary by as much as 1 magnitude
to indicate the possible scope of such an effect:
we find that our upper bound on $\Omega_{\Lambda}$ increases by about 0.06.

We find that using photometric redshifts from two different catalogs [6,7]
yields results that are almost indistinguishable.
Thus, we have shown that photometric redshifts 
can be used to estimate the expected number
of lensed galaxies in the HDF with reasonable accuracy.
This bodes well for the upcoming Southern Hubble Deep Field
redshift catalog that is expected in the near-future.
The Southern HDF will double the number of high redshift galaxies
and will increase the expected number of gravitationally lensed galaxies.
The actual number of lensed sources will lead to stronger constraints
on $\Omega_m-\Omega_{\Lambda}$.

\begin{figure}
\centerline{\epsfig{file=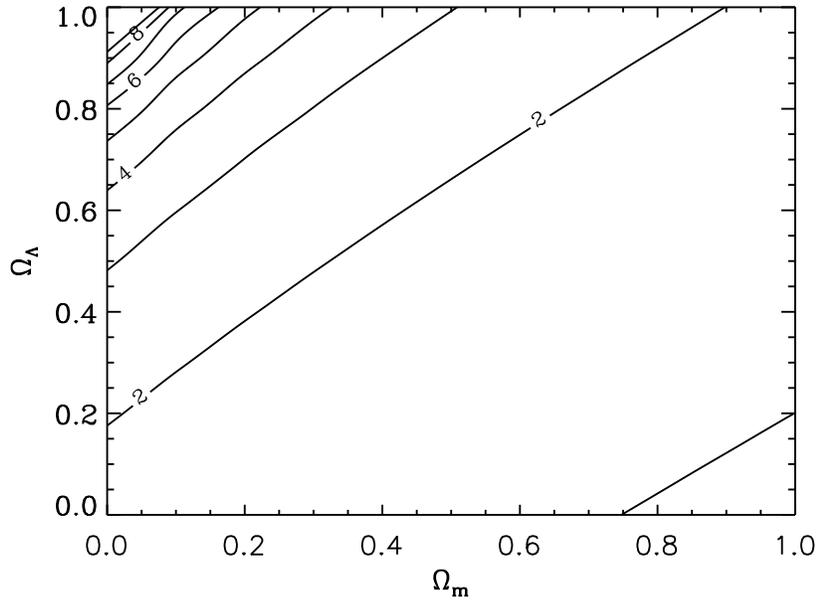,width=3.1truein,angle=90}}
\vspace{10pt}
\caption{Expected number of multiply-imaged galaxies, $\bar N$,
in the HDF, as a function of $\Omega_m$ and $\Omega_\Lambda$.
$\bar N$ is constant along lines of constant $\Omega_m-\Omega_\Lambda$,
allowing for direct constraints on this quantity.
Shown here is the expected number based on the redshift catalog of Ref. [6], 
and for lens search programs that have been carried out to an I-band
limiting magnitude of $m_{\rm lim}=28.5$.}
\end{figure}


\begin{references}

\bibitem{wil96} Williams, R. E., et al. 1996, AJ, 112, 1335.

\bibitem{hog96} Hogg, D. W., et al. 1996, ApJ, 467, L73.

\bibitem{zepf97} Zepf, S. E., et al. 1997, ApJ, 474, L1.

\bibitem{cor98} Cooray, A. R., Quashnock, J. M., \& Miller, M. C. 1999, 
ApJ, 511, in press.

\bibitem{fuk92} Fukugita, M., Futamase, T., Kasai, M., \& Turner, E. L. 1992, 
ApJ, 393, 3.

\bibitem{saw97} Sawicki, M. J., Lin, H., \& Yee, H. K. C. 1997, AJ, 113, 1.

\bibitem{wang98} Wang, Y., Bahcall, N., Turner, E. L. 1998, AJ, 116, 2081. 

\bibitem{rie98} Riess, A. G., et al. 1998, AJ, 116, 1009.

\bibitem{koc96} Kochanek, C. S. 1996, ApJ, 466, 638.

\bibitem{fac98} Falco, E. E., Kochanek, C. S., \& Munoz, J. A. 1998, 
ApJ, 494, 47.

\end{references}
\end{document}